# DSCEP: An Infrastructure for Distributed Semantic Complex Event Processing

Vitor Pinheiro de Almeida
Pontifícia Universidade Católica do Rio de Janeiro
Rio de Janeiro, Brazil
valmeida@inf.puc-rio.br

Markus Endler
Pontifícia Universidade Católica do Rio de Janeiro
Rio de Janeiro, Brazil
larst@affiliation.org

Sukanya Bhowmik
University of Stuttgart
Stuttgart, Germany
sukanya.bhowmik@ipvs.uni-stuttgart.de

Kurt Rothermel
University of Stuttgart
Stuttgart, Germany
kurt.rothermel@ipvs.uni-stuttgart.de

## ABSTRACT

Today most applications continuously produce information under the form of streams, due to the advent of the means of collecting data. Sensors and social networks collect an immense variety and volume of data, from different real-life situations and at a considerable velocity. Increasingly, applications require processing of heterogeneous data streams from different sources together with large background knowledge. To use only the information on the data stream is not enough for many use cases. Semantic Complex Event Processing (CEP) systems have evolved from the classical rule-based CEP systems, by integrating high-level knowledge representation and RDF stream processing using both the data stream and background static knowledge. Additionally, CEP approaches lack the capability to semantically interpret and analyze data, which Semantic CEP (SCEP) attempts to address. SCEP has several limitations; one of them is related to their high processing time. This paper provides a conceptual model and an implementation of an infrastructure for distributed SCEP, where each SCEP operator can process part of the data and send it to other SCEP operators in order to achieves some answer. We show that by splitting the RDF stream processing and the background knowledge using the concept of SCEP operators, it's possible to considerably reduce processing time.

## 1 INTRODUCTION

There are over 3.36 billion active smartphones today with an internet connection. In 2018 the number of worldwide mobile network users was 2.65 billion, and it is expected to grow to around 3.02 billion in 2021. Moreover, the average daily social media usage of internet users worldwide reached 136 minutes per day. These numbers indicate that almost a third of the entire global population is somehow generating information through their mobile phones [1].

Consequently, we are demanding more from data stream analysis systems that many times need to correlate data produced by smartphones or sensors in the form of streams with background knowledge in order to make sense of the data and with it achieve some conclusion [10, 20]. Increasingly, applications work with complex domains, which means that they require the processing of heterogeneous data streams together with background knowledge in order to make sense of the data stream produced by smartphones and/or sensors [13, 23].

Let us consider an example of a traffic prediction system that can predict how the traffic will be on the next hours (*S1*). This system must use two different sources of data, one is the data stream from the drivers' smartphone containing GPS data. The second one would be the map of the region that contains the relation among the streets and the direction that the cars can flow. With the GPS data it's possible to infer which street the driver is the driver's velocity. Moreover, this system can use also data from social networks like Twitter[1] to find an explanation for why street A is experiencing slow traffic. In this case Twitter would be a new source of data stream used by the system in order to find possible explanations on why there is slow traffic at street A. With the goal of proposing a suitable infrastructure model to process both data from the stream and data from background knowledge, one possibility is to think on how to combine Complex Event Processing (CEP) [19] with Semantic Web technologies [6]. The combination of these two areas created the term SCEP, which means Semantic Complex Event Processing.

CEP technologies have years of experience with how to provide timely answers for queries over the data stream. It is widely used in many domain areas such as processing of social network data [3] and applications for stock market shares. CEP already proved that it can work with high data throughput and volume and still deliver timely answers for its queries. Classical CEP technologies are not targeted at analyzing data from the stream together with background knowledge; the background knowledge use is optional. SCEP, on the other hand, comes as a ramification of CEP targeting these applications that must use background knowledge to make sense of the data on the stream.

SCEP has a set of requirements, and its goal is to use both CEP and Semantic Web technologies to achieve it [11]: (1) Volume: Social networks have billions of active users; Cities with thousands of sensors monitoring different types of information; (2) Velocity: Sensors can generate thousands of observations per minute; Social media users that produce, on average, 2.9 million posts per minute; (3) Timely answers: Answers should be generated within a specific time window, which depends on the application scenario and needs. In a patient monitoring application, a dangerous situation should be detected within minutes; (4) Complex Domains: Use cases that work with complex domains must use background knowledge to make sense of the data on the stream; (5) Data Heterogeneity: Each static and streaming data source normally have its own format. For example, each social media has its own data models and APIs. Web services to access weather

---

[1] Twitter: https://twitter.com/

data and databases available through the internet provide their data using different data formats.

One of the advantages of Semantic Web technologies is that it works well with the heterogeneity of data formats and with how to represent complex domains. Ontologies are a vital part of Semantic Web technologies, and they have been extensively used to model domain-specific knowledge of different domains [24]. They can represent data at the "semantic" level, which is not connected to data structures and implementation strategies. Thus, due to ontology data independence, ontologies are well suited for integrating heterogeneous data sources, enabling interoperability among different data streams and background knowledge.

Recent works on RDF[16] stream processing (RSP) are focusing on *Velocity*, *Volume* and *Data Heterogeneity* requirements [2, 4, 9, 15, 17]. Part of them is concerned on how to create an RSP engine to provide low processing time. Usually, the RSP engines which provides low processing time they only process data on the data stream without using an KB. A second smaller group aims more on creating infrastructures to parallelize RSP engines but disregarding the use of an KB. By not combining a background knowledge with the data on the stream, it is not possible to fulfill the *Complex Domain* requirement since these domains need background knowledge to make sense of the stream data.

SCEP engines, which supports SCEP languages[13], they are different from RSP engines. SCEP engines are more specialized and can be defined by the following: (1) Stream must be represented with a sequence of RDF triples each of them annotated with a timestamp; (2) Combine RDF stream data with a background knowledge base to deduce new information; (3) Enable stream reasoning; (4) Work with multiple RDF streams; (5) Provides window management operators for processing RDF streams; and (6) An output stream of one SCEP engine should be ready to be an input of another SCEP engine.

The research gap which this work is focusing on is how to provide *Timely Answers* for *Complex Domains* use cases, such as *S1*, which must use the background knowledge to make sense of the data on the stream. More specifically, we focus on requirements 3 (Timely answers) and 4 (Complex Domains).

The main contributions of this paper are briefly summarized as follows: (A) A system model for a distributed infrastructure for semantic complex event processing. Such infrastructure provides features to enable RDF stream processor engines to become SCEP engines; (B) An implementation for this distributed SCEP infrastructure model; and (C) Test and evaluation of the implemented infrastructure using CSPARQL as the engine for RDF stream processing. The tests include the relation between processing time and the size of the background knowledge base. Since the access to the background knowledge is costly, we show that using an distributed infrastructure

To the best of our knowledge, DSCEP is the first distributed infrastructure focusing on SCEP. We show that by dividing the RDF stream processing and the background knowledge using the concept of SCEP operators, it is possible to reduce processing time considerably.

This paper provides a conceptual model and an implementation of a distributed infrastructure for SCEP. It discusses each functionality that such infrastructure should provide in Section 2. In Section 3 we provide a implementation of the conceptual idea discussed in Section 2. We further present a set of tests and evaluations of the implemented infrastructure in Section 4. In Section 5, we give an overview of distributed RDF stream processing systems and infrastructures as well as a comparison between each of them with this work. Finally in Section 6 we conclude the paper with a brief discussion on the advantages and disadvantages of our proposed infrastructure and also talk about our next steps.

## 2 CONCEPTUAL ARCHITECTURE

In this section, we describe our conceptual model for a distributed infrastructure for SCEP. Our focus is on providing an infrastructure to distribute the RDF stream processing and to divide the KB through different machines to improve scalability and performance. The following are the assumptions made to this proposed infrastructure: (1) Every message sent through the infrastructure will always reach its destination; (2) Neither the machines nor the software required to run the infrastructure will fail; and (3) The timestamp of the RDF stream always increases. It means that an RDF stream processor will never receive an event with a timestamp older than the timestamp of the last processed event.

In summary, our infrastructure model has three modules. Figure 1a illustrates each of the modules: the *Stream Generator*, the *SCEP Operator*, and the *Client*. It is possible to have any number of each module running in the same infrastructure. Also, each module is independent and can run on the same or a different machine. The *Stream Generator* (Figure 1b) is the module which any *Script* capable of generating a RDF stream can be attached to. Since the stream is represented using RDF, the *Stream Generator* must offer two different forms to the *Script* (Figure 1b) for representing their stream events. The first is defining that an event is represented by a single RDF triple, for simpler data streams where each event is a single observation.

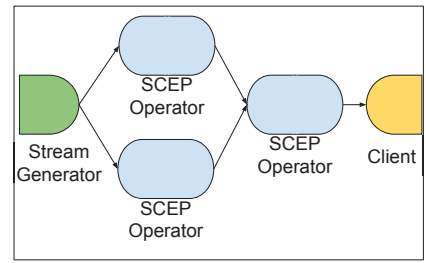

**(a) Infrastructure Conceptual Model.**

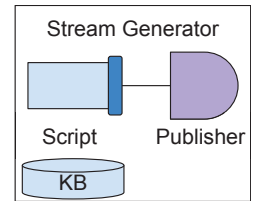

**(b) Stream Generator Module.**

**Figure 1: Infrastructure overview and Stream Generator Module.**

The second type of event on an RDF stream is the RDF graph, where a single event is composed of more than one RDF triple. The RDF graph event allows structuring more complex events in a stream, as opposed to plain triples. In our model, we define that each RDF triple in the RDF graph must contain a timestamp. This decision is because some RSP engines do not support RDF graph-based events, so they need that every triple contains its own timestamp.

The second module is the *SCEP operator* illustrated in figure 2a. This module is responsible for processing the input RDF stream and for generating an output stream as a result. This module is divided into *Aggregator*, *RSP engine* and *Publisher* (see Figure 2a). All *SCEP operator*'s modules do not need to be running on the same machine.

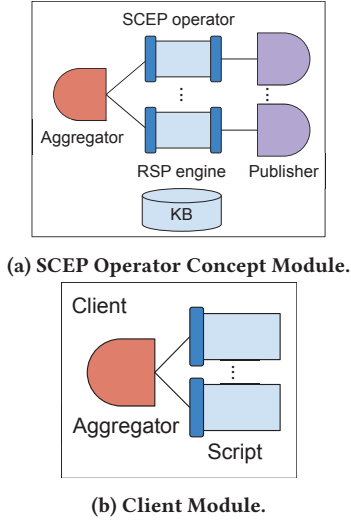

(a) SCEP Operator Concept Module.

(b) Client Module.

Figure 2: SCEP Operator and Client Modules.

The *Aggregator* (Figure 2a) is responsible for receiving all the input RDF streams from others *SCEP operators* or from a *Stream Generator*. The *Aggregator* will merge all input RDF streams into one, order the events on the new resulting stream, divide it into windows and send it to the attached *RSP engine*. As the *Aggregator* is responsible for managing the windows, it is possible to attach more than one *RSP engine* to it and send each window to a different *RSP engine*. By sending the windows to different *RSP engines*, it is possible to improve the parallelization and scalability of the infrastructure further. The *Aggregator* will enable the *RSP engine* attached to it to work with different types of windows and also to accept multiple streams. Such a feature is important because most RSP engines' implementation does not fill the requirements to work as a SCEP engine.

The *RSP engine* (Figure 2a) is responsible for processing each window sent by the *Aggregator* and to produce a RDF stream as an output. The user of the infrastructure can choose which RSP engine to use. The requirement is that the RSP engine can process RDF streams and produce RDF triples as output. The *Publisher* (Figure 2a), is the last part of the *SCEP operator*, and it is responsible for receiving the resulting RDF stream from it's respective *RSP engine*. Some RSP engines' implementation does not include, on their output RDF stream, the timestamp of each RDF triple. As a consequence, the *Publisher* can add a timestamp on each RDF triple if they do not have one. Moreover, in case that the stream's event is represented using an RDF graph, the *Publisher* is also responsible for identifying which set of RDF triples corresponds to an RDF graph.

The last module of the SCEP infrastructure is the *Client* module (figure 2b), where an end-user can attach his/her *Script* and make use of the RDF stream data. The *Client* module's *Aggregator* will receive all RDF streams which this *Client* is interested in, merge them, order them and send to the attached *Script*. The *Script* is where the user of the infrastructure can add their code to make use of the RDF stream data. It is possible for the user to create multiple *Scripts*, and this will make the infrastructure to balance the load among them. Thus, each window from the data stream will be sent by the *Aggregator* to the available *Script*.

To summarise, our conceptual model for a distributed infrastructure for SCEP enables two different kinds of query execution parallelism: *inter-query* parallelism and *intra-query* parallelism [26]. *Inter-query* parallelism, is when different queries execute in different *SCEP operators*. Each *SCEP operator* will have their own set of *RSP engines* and can run in different machines.

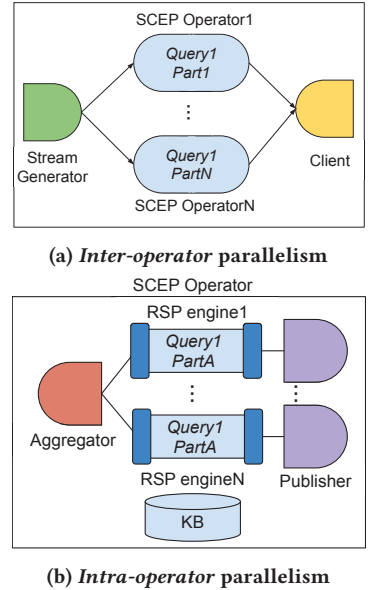

(a) *Inter-operator* parallelism

(b) *Intra-operator* parallelism

Figure 3: *Intra-query* parallelism

The *intra-query* parallelism concerns the parallelization of one query within different SCEP operators. *Intra-query* parallelism is further divided into *inter-operator* and *intra-operator* parallelism (Figure 3). *Inter-operator* parallelism, illustrated on Figure 3a, is when one query is divided into sub queries where each sub query executes in a different *SCEP operator*. Every sub query will receive the same data stream, but will execute in different *SCEP operators*. The *intra-operator* parallelism, illustrated on Figure 3b, refers to executing the same sub query at different *RSP engines* of the same *SCEP operator*. Since every *RSP engine* will have the same sub query, the *Aggregator* will divide the windows among the *RSP engines*. By doing so, the load of one stream will be divided into multiple *RSP engines*. Each *RSP engine* can run in a different machine.

## 3 IMPLEMENTATION

This section describes DSCEP, which is our implementation of the proposed conceptual infrastructure for distributed SCEP presented in section 2. The communication among the *Stream Generator*, *SCEP Operator* and *Client* modules is implemented using Apache Kafka[2]. Apache Kafka is a distributed streaming platform that enables its users to publish and subscribe to data streams. Additionally, Apache Kafka stores the data streams in a fault-tolerant way guaranteeing no message loss.

---

[2]Apache Kafka: https://kafka.apache.org/

DSCEP is implemented using Java and uses two different APIs from Apache Kafka, the Consumer API, and the Producer API. The Producer API allows the user application to publish a data stream to one or more Kafka topics while the Consumer API allows the user application to subscribe to a Kafka topic in order to receive a data stream.

The *Stream Generator* (Figure 1b) and *SCEP Operator* (Figure 2a) modules uses the Kafka's Producer API on its *Publisher* part. The *Publisher* is a Kafka Producer that must be created by the user to publish the data stream to a Kafka's topic. The user can use any programming language that supports Kafka. Additionally, the user has to create a name for the Kafka's topic to publish the data stream. The data stream also has to be published according to one of the message formats supported by DSCEP. The message formats are represented in JSON, and there is one for data streams of RDF triples and one for data streams of RDF graphs.

The *SCEP Operator* (Figure 2a) and the *Client* (Figure 2b) uses Kafka's Consumer API on it's *Aggregator*, *RSP engine* and *Script* parts. All *RSP engines* from the same *SCEP Operator* are part of the same consumer group. The consumer group functionality is available through Kafka's Consumer API. All consumers from the same consumer group process a different event from each other. This characteristic implies that every consumer will never process an event that another consumer on its group already processed. The *Aggregator* will publish the data stream in the form of windows. Whenever a connected *RSP engine* is available, it will take one window from the published data stream to process. All *Scripts* from the same *Client* module are also from the same consumer group.

All *RSP engines* will subscribe to its related *Aggregator* topic. Every *RSP engine* must have only one *Aggregator* associated to it. For the user to select which data streams to receive, there is a file named *node.properties*, located in the same folder where the *Aggregator* is running. This file contains the attribute "topics", where the user can write all topics that the *Aggregator* must subscribe to. Finally, its possible to attach any new *Stream Generator*, *SCEP operator* or *Client* while the DSCEP is running. If a *Client* or a *SCEP Operator* module is attached while DSCEP is already running, they will not receive any past events from data streams since DSCEP does not persist them.

## 4 EVALUATION

### 4.1 Datasets

For the example use case, we used the TweetsKB[3] dataset to simulate the data stream. TweetsKB [12] is a public RDF corpus of anonymized data for an extensive collection of annotated tweets. The dataset currently contains data for more than 1.5 billion tweets, spanning more than five years. Metadata information about each tweet is available using well established RDF vocabularies. To transform this dataset into a data stream, we made each tweet an RDF graph, and we inserted a timestamp into each RDF triple. The timestamp of each RDF triple of a tweet is the timestamp of the tweet's creation.

All tweets on the dataset contain different types of information. On the following, we explain all types of information of a tweet that we used on our use case: (A) *TweetID*: A unique number that identifies the tweet on the whole dataset; (B) *Entities*: The entities are extracted from the tweet's text. Each tweet can contain any number of entities. Each entity is related to one resource in DBpedia that describes the entity; (C) *Sentiment Analysis*: Is a positive or a negative score applied to the tweet. The number of each sentiment (positive or negative) ranges from 0.0 to 5.0; and (D) *Likes and Shares*: Its the number of likes and shares of the tweet. Each tweet can have any number of likes and shares.

For the tests, we used one month of data of TweetsKB that contains approximately 60 thousand tweets that correspond to a total of 2,3 million triples. For the background knowledge, we use the DBpedia dataset [7]. DBpedia is an RDF KB built by a community effort that extracted structured information from Wikipedia, making this information accessible on the Web. Currently, DBpedia has a total of approximately 370 million RDF triples. We choose DBpedia because each tweet of the TweetKB is already related to an entity URI on DBpedia. Thus, it is possible to enrich each tweet information using DBpedia data. The DBpedia dataset size used is the same as the public available DBpedia endpoint[4].

### 4.2 Evaluation Setup

DSCEP was tested and deployed on a machine with 512 GB of RAM and two AMD EPYC 7451 processors. Each processor has 24 cores with 2.3GHz and a 64MB of cache. For the communication among the nodes of the infrastructure, we used Apache Kafka version 2.0 and Zookeeper[14]. To facilitate the deployment of DSCEP modules, we used Docker. Each DSCEP module (*Stream Generator*, *SCEP Operator* and *Client*) runs in a different Docker container. Additionally, we run our own DBpedia endpoint using Virtuoso[5] on a separately docker container. All SCEP Operators containers have access to the DBpedia endpoint container. We choose C-SPARQL as the RDF stream processor of each *SCEP Operator*.

### 4.3 Method

To evaluate how DSCEP can contribute to decreasing processing time of continuous queries on RDF streams, we first use the SR-Bench benchmark[25]. SR-Bench classifies continuous SPARQL queries into different types of queries. Our focus is on systems that need to query both the stream and the KB.

*First step:* In our first step, we adapted queries Q15 and Q16 from SR-Bench to test with C-SPARQL within DSCEP. These queries exploit the RDF processor ability to apply reasoning using properties *rdfs:subClassOf* and *owl:sameAs*. *Q15* contains hierarchy reasoning, and *Q16* contains a property path expression. Query *Q15* gets all tweets that mention any entity that is a subclass of MusicalArtist. Query *Q16*, for every tweet that has an entity of type Musical Artist, it will also return the birthplace, country and country code.

These are simple queries to test how the RDF processor behaves with different characteristics of continuous SPARQL. This first step will give us the average processing time of C-SPARQL for basic queries with TweetKB as the data stream and DBpedia as the background KB.

For all queries made during the evaluation, we tested them using two different methods of accessing the KB. The first method is using **C-SPARQL KB access** method. This method is used by C-SPARQL to include an RDF file as the background knowledge for every window. The second method we use C-SPARQL but with a **SPARQL sub query** (using the SERVICE operator of the SPARQL language) to access the background KB data. For the subquery method, the KB is located in a different docker container.

---
[3]TweetsKB: http://l3s.de/tweetsKB/
[4]DBpedia: https://wiki.dbpedia.org/public-sparql-endpoint
[5]Virtuoso: https://virtuoso.openlinksw.com/

*Second step:* The second step is focused on evaluating a more complex query that could be used in a real-world scenario and show how DSCEP can decrease query evaluation time. The goal is to compare the processing time of one complex query when executed with one C-SPARQL query, with when executed in parallel by dividing it into multiple subqueries. The query used in this step, which we will name by *CQuery1*, contains the following SPARQL characteristics:

- Property Path expressions: KB and data stream are interlinked; the maximum path length is 3.
- Construct derived knowledge: Necessary to build an output RDF stream.
- Union: *FILTER* operator of SPARQL.
- Optional Pattern Matching: *OPTION* operator of SPARQL.
- Ontology-Based Reasoning: Hierarchical reasoning using *rdfs:subclass*.
- KB access: Requires access to the KB.

*CQuery1* objective is to evaluate how the sentiment analysis of entities of the class *MusicalArtists* are affected when mentioned on the same tweet with entities of the class *TelevisionShows*. In other words, how television show entities affect the sentiment analysis of each musical artist when they are mentioned on the same tweet. For example, let us assume that artist *Bob*, when mentioned with a television show (*ShowA*), always has a positive sentiment analysis. It means that the probability of *Bob* of being positively mentioned when related in the same tweet of *ShowA* is 100%.

*CQuery1* graph when divided into subqueries is illustrated on Figure 4. Each blue box in the figure runs in a different docker container, and the DBpedia KB is also located in a separate docker container. Each subquery is located in a different SCEP Operator. *QueryA* and *QueryB* are executed in parallel, *QueryC*, *QueryD*, *QueryE* and *QueryF* are also executed in parallel. All of them use inter-query parallelism. *QueryG* only aggregates the resulting streams and correlates how musical artists are associated with television shows.

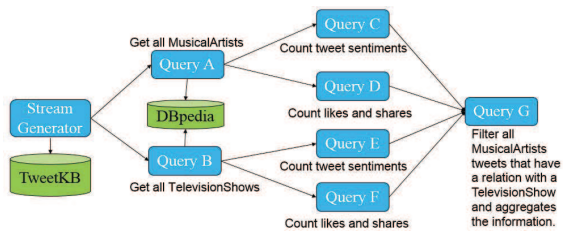

**Figure 4:** *CQuery1* **graph when divided into sub queries.**

All results are the same when executing *CQuery1* with only one C-SPARQL and when dividing it, as illustrated in Figure 4. The second step shows how dividing a query into subqueries using DSCEP can decrease processing time, using two different forms of accessing the KB.

*Third step:* For further evaluation, in the third step, we compare how the KB total size and the KB used size can affect the processing time of the subqueries used in the second step. The total KB size is the number of RDF triples in the KB. The used KB size is the actual number of RDF triples used by the query. The third step helps us to understand how processing time can increase depending on the KB size.

## 4.4 Results and Discussion

*Results of the First step:* During the first step, the throughput is 50.000 RDF triples per second, and the window size is a maximum of 1000 RDF triples. The events are made of RDF graphs, but the window size is calculated in the number of triples. Thus, DSCEP aggregates as many RDF graphs that their sum of triples is a maximum of 1000 RDF triples.

Table 1 shows that the **C-SPARQL KB access** method has a lower processing time when executing queries with Property Path expressions compared with using the **SPARQL subquery** method. On the other hand, with *Q15*, which uses hierarchical reasoning (*rdfs:subclass*), the **SPARQL subquery** method demonstrates lower processing time than the **C-SPARQL KB access** method. These numbers are important for us to have an idea of how many seconds C-SPARQL takes to execute basic queries that require KB access.

*Results of the Second step:* During the second step, the throughput is 25.000 RDF triples per second, and the window size is a maximum of 1000 RDF triples. The events are made of RDF graphs, but the window size is calculated in the number of triples. Thus, DSCEP aggregates as many RDF graphs that their sum of triples is a maximum of 1000 RDF triples.

Table 2 shows the results for when *CQuery1* is executed as one C-SPARQL and Table 3 shows the results for when *CQuery1* is executed in parallel (according to Figure 4). With the **C-SPARQL KB access** method, the results of the second step show us a reduction on the query processing time of 29% when executing *CQuery1* in parallel (Table 3) compared to run it all in one query (Table 2). When using the **SPARQL subquery** method, the reduction on query processing time is of 23% when executing *CQuery1* in parallel compared to run it all in one query. Additionally, the processing time elapsed by queries *QueryC*, *QueryD*, *QueryE*, *QueryF* and *QueryG* are a total of 36,2 ms. Since these queries does not access KB, their processing time are much lower.

This result demonstrates that by dividing a query into subqueries that can be parallelizable, it is possible to reduce processing time. Also, these tests show that the KB access is costly. The third step of the results will show us how the KB can affect the processing time.

*Results of the Third step:* First, we demonstrate how variating the number of used triples in the KB can affect the processing time, without changing the total KB size. Figure 5a shows that when used KB size is 103075, it takes 81,34 seconds to process. Although, if the used KB size is only 10401, it takes 8,41 seconds to process. This result shows that when the used KB size is reduced approximately by ten times, the processing time also reduces approximately by ten times. It concludes that if we divide one query into subqueries that use smaller parts of the KB, it is possible to reduce processing time significantly.

Similar behavior happens when testing the variation of used KB triples with *QueryB*, illustrated by Figure 5b. When reducing the used KB size by approximately 7,5 times (from 29414 triples to 3994 triples), the processing time also reduces by approximately 6,5 times (from 9,56 secs to 1,53 secs). The reduce factor in processing time is lower than with the test of *QueryA*, showing that as smaller the number of used triples is, the less it affects the processing time. Figures 6a and 6b shows how the unused triples in the KB can affect the processing time. Both figures are marked with a circle on value 8,57 secs; this is the processing time of QueryA when used KB size is equal to the total KB size. Figure 6b shows that when the total KB size increases from 10401 to 103075 triples, the processing time increases by 30,2%. This increase in processing time is directly related to the number of unutilized triples by *QueryA* on the KB. Figures 7a and 7b demonstrates that *QueryB* has a similar effect in processing time when increasing total KB size. When total KB size increases from 3994 to 29414 triples, the processing time increases approximately 43,6% (from 2,8 seconds to 4,02 seconds). It shows that the smaller the KB size is, the processing time can decrease exponentially.

Table 1: Results of the first step.

|  | C-SPARQL KB access | | SPARQL subquery | |
|---|---|---|---|---|
|  | *Q15* | *Q16* | *Q15* | *Q16* |
| Total KB size (Nº of triples) | 103.075 | | 368.720.213 | |
| Used KB size (Nº of triples) | 103.075 | | 103.075 | |
| Processing time (per window) | 5 secs | 0,64 secs | 1,3 secs | 1,61 secs |

Table 2: Results of the second step: *CQuery1* as one C-SPARQL query.

|  | *CQuery1* | |
|---|---|---|
|  | C-SPARQL KB access | SPARQL subquery |
| Total KB size (Nº of triples) | 132.489 | 368.720.213 |
| Used KB size (Nº of triples) | 132.489 | 132.489 |
| Processing time (per window) | 117,05 secs | 104,35 secs |

Table 3: Results of the second step: *CQuery1* divided into multiple C-SPARQL queries (see Figure 4).

|  | C-SPARQL KB access | | SPARQL subquery | |
|---|---|---|---|---|
|  | *QueryA* | *QueryB* | *QueryA* | *QueryB* |
| Total KB size (Nº of triples) | 103.075 | 29.414 | 368.720.213 | |
| Used KB size (Nº of triples) | 103.075 | 29.414 | 103.075 | 29.414 |
| Processing time (per window) | 84,66 secs | 26,65 secs | 81,33 secs | 22,82 secs |

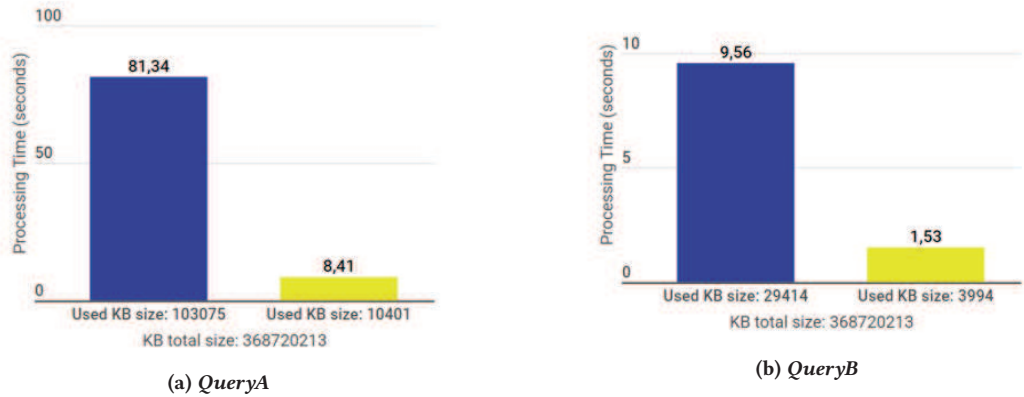

(a) *QueryA*

(b) *QueryB*

Figure 5: Variating used KB size. (SPARQL subquery method)

## 5 RELATED WORKS

Several RSP engines have been developed in the last decade, some focusing on the processing aspects of continuous RDF streams and other focusing on providing query expressiveness and reasoning capabilities [2, 4, 9, 15, 17]. CQELS-cloud [18] was the first, which mainly focuses on the engine scalability and elasticity. The CQELS-cloud main contribution was its query analyzer, which parallelizes different aspects of the query into Apache Strom[6] to decrease query processing time. However, CQELS-cloud focus on parallelizing a single query execution and not to provide an infrastructure for connecting multiple queries.

Calbimonte, in [8], proposed the first distributed infrastructure for RDF stream processing, focusing on connecting different RSP engines. The infrastructure allows the developer to plug in an RSP engine or an RDF stream generator and use the infrastructure to make them communicate. The infrastructure provides communication among its nodes using AKKa HTTP. The focus of Calbimonte's work is on how to connect different types of RSP engines, proposing the use of web standards to enable RSP engines' developers to plug in their RDF processors.

One difference between Calbimonte's infrastructure and DSCEP, is that DSCEP is focused on working with SCEP. DSCEP incorporates features that can enable an RSP engine to work as a SCEP engine. For example, Calbimonte's infrastructure has

---
[6]Apache Strom: https://storm.apache.org/

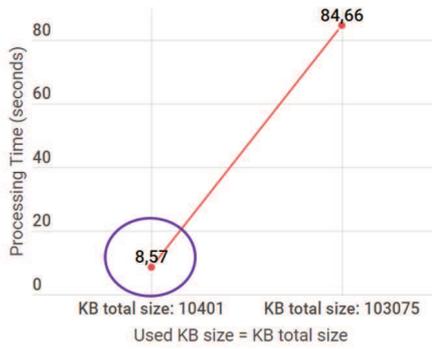
(a) Used KB size = Total KB size.

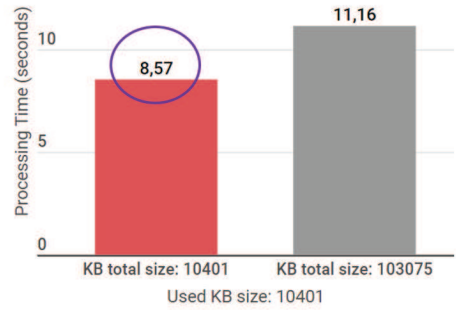
(b) Used KB size < Total KB size.

Figure 6: *QueryA* using C-SPARQL KB access method

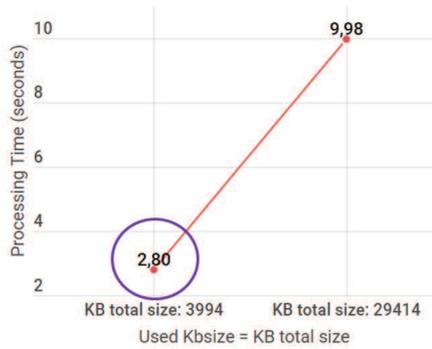
(a) Used KB size = Total KB size.

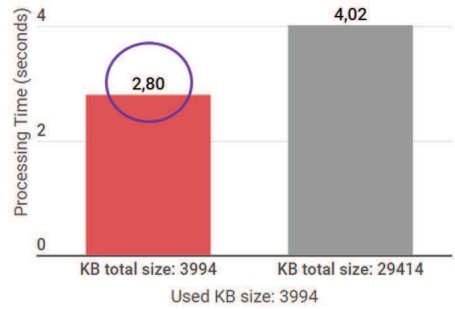
(b) Used KB size < Total KB size.

Figure 7: *QueryB* using C-SPARQL KB access method

restrictions for RSP engines to access a database; they can be only locally accessed. DSCEP also offers features such as window management and stream aggregation, which are required characteristics for an RSP engine to be considered a SCEP engine. Moreover, DSCEP also provides two different types of query execution parallelism: *inter-query* parallelism and *intra-query* parallelism. Calbimonte's infrastructure can only parallelize using *inter-query* parallelism method.

Xiangnan Ren in [21] proposed Strider, which is also a distributed infrastructure for RDF stream processing. Different from DSCEP, Strider's infrastructure uses its own query language on its RDF stream processor. Strider's query is transformed into Spark Streaming[7] queries to enhance parallelization capabilities. Additionally, Strider's infrastructure does not support access to KBs during query execution and does not provide reasoning capability. Therefore, Strider is not an infrastructure ready for distributed SCEP.

After Strider, Xiangnan Ren in [22] proposed BigSR, which is an improvement of the Strider infrastructure. The improvements are on enabling recursive and more expressive queries. To enhance expressiveness, BigSR uses LARs [5] as its query language, which can include logical axioms and logical rules. This improvement enables BigSR to execute reasoning with the logical rules included in the LARs query. BigSR still does not offer access to external background knowledge on the query level.

Table 4 gives a comparison among the infrastructures mentioned so far and it summarize their differences. The following is the explanation of each comparison made on table 4.

- *Query execution parallelism types*: State whether the infrastructure accepts *inter-query* parallelism or *intra-query* parallelism or both.
- *Accept different RSP engines*: State whether or not the infrastructure can be used with different RSP engines. This feature is important because there are not many SCEP engines ready to be used. An infrastructure for SCEP must enable an RSP engine to work as a SCEP engine.
- *Allow KB access*: State whether the infrastructure allows RDF engines to access external or local KBs on query level.
- *Enable reasoning (stream+KB)*: State whether the infrastructure enables RSP engines to do reasoning using data both from the stream and from a knowledge base.
- *Built for SCEP*: State whether the infrastructure focus is to support SCEP engines.

From table 4, it is possible to conclude that, over time, KB access is becoming more restricted or even nonexistent. The main reason for that is because the infrastructures are more concern with decreasing processing time and latency. To access the KB is more time costly than to only process data on the stream.

DSCEP's focus is to be a distributed infrastructure for SCEP; thus, it is essential to enable access to knowledge bases. In section 4, we provide experiments and show how the knowledge base can affect the processing time of queries. To provide a distributed

---
[7]Spark Streaming: https://spark.apache.org/streaming/

Table 4: Comparison among distributed infrastructures for RDF stream processing

|  | Query execution parallelism types | Accept different RSP engines | Allow KB access | Enable Reasoning (stream+KB) | Built for SCEP |
|---|---|---|---|---|---|
| D. L. Phuoc, 2013 | Inter-query | ✗ | ✗ | ✗ | ✗ |
| Calbimonte, 2017 | Inter-query | ✓ | ✓* | ✓ | ✗ |
| Xiangnan, 2017 | Inter-query Intra-query | ✗ | ✗ | ✗ | ✗ |
| Xiangnan, 2018 | Inter-query Intra-query | ✗ | ✗ | ✓ | ✗ |
| DSCEP | Inter-query Intra-query | ✓ | ✓ | ✓ | ✓ |

* KB is only locally accessible.

infrastructure for RDF processors which their queries must access a KB, is one step towards reducing query processing time. Processing time can be reduced by parallelizing query execution and by dividing the KB through multiple machines.

## 6 CONCLUSION

In this paper, we present DSCEP, which, at best at our knowledge, is the first distributed infrastructure focused on Semantic CEP. We show that by dividing a query into subqueries and parallelize them with DSCEP infrastructure, it is possible to reduce processing time up to 29% without changing the query results. Moreover, we show how accessing a knowledge base can affect the processing time of an RSP engine. Our tests demonstrate that by dividing the KB in a way that each subquery only accesses its own part of the KB, it is possible to decrease processing time. Additionally, DSCEP provides features to enable a variety of RSP engines to work as SCEP engines. Features such as window management, support to streams of RDF graphs, enable RSP engines to work with multiple streams and offers two types of parallelism (inter-query and intra-query).

Research must be done within the area of SCEP engines. The scalability of current RSP engines when combining data from the stream with a knowledge base is still an issue. It can be mitigated by proposing new RDF processing algorithms and new distribute infrastructures.

As future work, we plan to enable DSCEP to offer the possibility of dividing the KB automatically among its operators. Since all queries are predefined, DSCEP can identify the part of the KB that each SCEP operator needs. By doing so, DSCEP can send each part of the KB to its respective SCEP operator to decrease processing time. Another future work is to test our infrastructure with other RSP engines and also enable DSCEP to execute operator and database placement. During runtime, it is possible to read different parameters if the infrastructure and test different positions within the infrastructure to place operators and databases in order to decrease processing time.